\documentclass[twocolumn,showpacs,aps,eqsecnum,pra]{revtex4-1}

\usepackage{amsmath}
\usepackage{color}
\usepackage{amssymb}
\usepackage{epsfig}
\usepackage{soul}
\usepackage{bbm}
\usepackage{bm}
\usepackage{bbold}
\DeclareMathOperator{\sgn}{sgn} 
\begin{document}
%\pagenumbering{arabic}
\title{ Einstein-Podolsky-Rosen uncertainty limits for bipartite multimode states} 
\author{ Paulina Marian}
\email{paulina.marian@g.unibuc.ro}
\author{ Tudor A. Marian}
\email{tudor.marian@g.unibuc.ro}
\affiliation{ Centre for Advanced  Quantum Physics, Department of Physics, University of Bucharest,
R-077125 M\u{a}gurele, Romania}

%\date{\today}

\begin{abstract}
 
Certification and quantification of correlations for multipartite states of quantum systems 
appear to be a central task in quantum information theory.  We give here a unitary 
quantum-mechanical perspective of both entanglement and Einstein-Podolsky-Rosen 
(EPR) steering of continuous-variable multimode states. This originates in the  
Heisenberg uncertainty relations for the canonical quadrature operators of the modes. 
Correlations of two-party $(N\, \text{vs} \,1)$-mode states are examined by using 
the variances of a pair of suitable EPR-like observables. It turns out that 
the uncertainty sum of these nonlocal variables is bounded from below by local 
uncertainties and is strengthened differently for separable states and for each 
one-way unsteerable ones. The analysis of the minimal properly normalized 
sums of these variances yields necessary conditions of separability and EPR
unsteerability of $(N\, \text{vs} \,1)$-mode states in both possible ways of steering. 
When the states and the performed measurements are Gaussian, then these conditions 
are precisely the previously-known criteria of separability and one-way unsteerability. 
  
\end{abstract}

\maketitle

\section{Introduction}

The use of uncertainty arguments to understand quantum properties of composite systems 
was initiated in the seminal work of Einstein, Podolsky and Rosen \cite{EPR} and deepened
by Schr\"odinger, shortly thereafter  \cite{Sch}. Specifically, in Refs.\cite{EPR,Sch} 
the following paradox of quantum mechanics is described: two distant parties of a given system 
share an entangled state and one party, by measuring its subsystem, can remotely change (steer) 
the state of the other party's subsystem, in contrast to the requirements of the local realism. 
Many years after, Bell proved that no local realistic theory can give a complete description 
of the predictions of quantum mechanics (Bell's theorem) \cite{Bell}.  A further contribution 
in defining quantum correlations between the parties of a composite system  was made
by Werner \cite{Werner}.  What was subsequently known as a separable state, namely, 
a convex combination of product states, was termed by Werner as being classically 
correlated. Werner's original terminology for all the states not having such an expansion 
has been Einstein-Podolsky-Rosen (EPR) correlated states. Nowadays they are called entangled.  
Soon after Werner's definition, separability conditions were found for finite-dimensional 
systems in terms of $2$-entropy inequalities \cite{H1,H2} or preservation of the non-negativity 
of the density matrix under partial transposition \cite{Peres,H3}. Note that in 1989, a practical 
procedure to demonstrate the EPR paradox by using  products of inferred variances 
was first proposed by Reid \cite{Reid1} for continuous-variable bipartite states. 

A significant progress in understanding quantum correlations manifested 
through entanglement, EPR steering, and Bell non-locality, in the framework of a unified 
quantum-information description was given by Wiseman {\em et al.} as three rather different
tasks for confirming inseparability \cite{W1,W2,W3}. First, for certification of  two-party entanglement, 
use is made of the general aspect of a separable state \cite{Werner} and quantum state tomography 
is performed with all-trusted measurement devices. Second,  quantum steering corresponds 
to the task of verifiable entanglement distribution by one untrusted party. Third, in the case 
of Bell nonlocality, both parties do not trust their measuring devices. In the EPR steering scenario, 
the two parties, Alice and Bob, play a different role and thus EPR steering is asymmetric 
when interchanging the subsystems, unlike entanglement and Bell nonlocality \cite{W2}. 
Being perceived as another type of quantum correlations, EPR steering has recently attracted 
considerable interest especially related to its verification and quantification from a quantum information 
perspective \cite{RMP,Pusey,SNC,QVB,UMG,PW,Adesso2015,Kiukas,exp2012,Das2019}. 
Note that for continuous-variable settings, much attention was given from the very beginning  
to the conditions and quantifiers of steering for Gaussian states (GSs)
\cite{Reid1,W2,Reid2,Adesso2,Adesso3,Nha,Kiukas2,He}. 

In the present paper  we take a different approach that treats on an equal footing
entanglement and steering. This generalizes and enlarges the EPR-line of reasoning 
initiated by Reid for a two-mode state \cite{Reid1}.  Essentially, Reid used the variances 
of two nonlocal observables linearly built with the one-mode canonical quadrature operators 
${\hat q}_j, {\hat p}_j, (j=1,2)$:
\begin{align}
{\hat Q}(\lambda):= {\hat q}_1-\lambda {\hat q}_2,       \qquad      
{\hat P}(\mu):= {\hat p}_1+\mu {\hat p}_2, 
\label{lm}
\end{align}
where $\lambda$ and $\mu$ are adjustable positive parameters. The coordinates 
and momenta in Eq. (\ref{lm}) are defined in terms of the amplitude operators of the modes:
\begin{align}
& \hat q_j:=\frac{1}{\sqrt{2}}( \hat a_j+\hat a_j^{\dag}),      \qquad
\hat p_j:=\frac{1}{\sqrt{2}i}( \hat a_j-\hat a_j^{\dag}).
\label{qp}
\end{align}
Unless  $\lambda \mu=1$, the operators (\ref{lm}) are not proper EPR observables
since they do not commute.
Consequently, we get the weak (Heisenberg) form of the uncertainty relation (UR)
\begin{equation}
\Delta Q(\lambda) \, \Delta P(\mu) \geqq \frac{1}{2} |1-\lambda \mu|, 
\label{prodUR} 
\end{equation}
which has to be fulfilled by any quantum state. 
In Eq. (\ref{prodUR}) and in the sequel as well, $\Delta A$ denotes 
the standard deviation of the observable $\hat{A}$ in the state ${\hat \rho}$, 
which is the square root of the variance
\begin{align}
\left( \Delta A \right)^2:= \left< \left( {\hat A}
-{\langle {\hat A }\rangle }{\hat I} \right)^2 \right> 
={\langle {\hat A}^2 \rangle }-{\langle {\hat A} \rangle}^2. 
\label{varA} 
\end{align}
In Refs. \cite{Reid1,Reid2}, a possible experimental observation of the inequality 
\begin{equation}
\Delta Q(\lambda) \, \Delta P(\mu) < \frac{1}{2}
\label{EPR}
\end{equation} 
was interpreted as a signature of detecting an EPR paradox. Moreover, 
the EPR paradox as invoked by Reid and the concept of EPR steering  
as analyzed in Refs. \cite{W1,W2,W3}  were proven to be equivalent.  
The way this equivalence arises from two different perspectives became a first reason 
for our interest in understanding the modification of the URs for unsteerable states. 
A second motivation for the present work stems from a recent result on EPR-like URs 
valid for separable two-mode states \cite{PT2018}.  Essentially, in Ref. \cite{PT2018}
it was proven that the minimum normalized product and sum of the uncertainties
of appropriate EPR-like observables are separability indicators for two-mode GSs, 
due to the Peres-Simon separability theorem. According to this, such a state is 
separable if and only if its density matrix has a positive-semidefinite partial transpose 
(PPT) \cite{Peres,Simon}.     
 
Consequently, the remainder of this paper develops an approach for a parallel study
of the separability and unsteerability of two-party $(N\, \text{vs} \,1)$-multimode states.
This study exploits the sum-form URs for a pair of appropriate EPR-like observables.
In Sec. II we extend the Reid treatment of two-mode states to the larger class 
of two-party $(N\, \text{vs} \,1)$-mode states. We thus define two linear combinations 
of canonical quadrature operators of the same kind, each one depending on
$N+1$ positive parameters, and then write their URs in an arbitrary bipartite 
$(N\, \text{vs} \,1)$-mode state. In Sec. III we apply an important theorem 
of Hofmann and Takeuchi \cite{HT} to find out the enhancement of the URs written 
for a separable $(N\, \text{vs} \,1)$-mode state. Accordingly, a modification 
of the lower bound of URs for separable states is imposed by observing 
the local uncertainty relations (LURs). By extremization of the normalized  
uncertainty sum with respect to the parameters, we then establish a necessary 
condition of separability which appears to reduce to the Peres PPT one. 
Section IV is devoted to an analysis of the steering process which aims to enable us
to put forward the efficient way of applying the Hofmann-Takeuchi theorem in the case
of one-way unsteerability. In Sec. V we perform the extremization of the properly
normalized uncertainty sums in order to get necessary conditions of unsteerability
for both ways of potential steering. It appears that they coincide with those found 
by Wiseman {\em et al.} \cite{W1,W2} in the Gaussian framework.  We revisit 
in Sec. VI the special case of GSs, for which it is well known that the necessary 
conditions of separability and unsteerability we have derived here are also sufficient 
ones, as proven in Refs. \cite {Simon, WW2001} and, respectively, 
in Refs. \cite{W1,W2}. Section VII contains a recapitulation and a discussion 
of the methods and results of the paper.

\section{EPR-like uncertainty relations for bipartite 
${\bm (}{\bm N} \, \text{vs} \, {\bm 1}{\bm )}$-mode states}
 
Qualification and quantification of quantum correlations for multimode states 
continue to be a central task of continuous-variable information theory 
\cite{LF2003,H2009,RMP}.  For simplicity and feasibility reasons, we consider here  
a two-party state ${\hat \rho}$ with $(N\, \text{vs} \,1)$ modes shared by Alice and Bob as follows. 
Alice performs measurements on the $N$-mode part of the state with the canonical 
quadrature operators ${\hat q}_j, {\hat p}_j, ( j=1,\dots, N)$,  while the  $(N+1)$th mode 
with the pair of canonical quadrature  operators $\hat q_{N+1}, \hat p_{N+1}$ is controlled by Bob. 
In order to generalize Eq.(\ref{lm}), we introduce the following  EPR-like observables depending 
on the positive parameters ${\alpha}_j >0, \;  {\beta}_j >0, \\  (j=1,\dots, N+1)$:  
\begin{align} 
& {\hat Q}({\bm \alpha}):=\sum_{j=1}^{N}{\alpha}_j {\hat q}_j-{\alpha}_{N+1} {\hat q}_{N+1},   
\notag  \\
& {\hat P}_{\pm}({\bm \beta}):=\sum_{j=1}^{N}{\beta}_j {\hat p}_j \pm {\beta}_{N+1} {\hat p}_{N+1},           
\quad  \text{where}     \notag   \\
& {\bm \alpha}:=\{ {\alpha}_1, \dots , {\alpha}_{N+1} \}, 
\qquad  {\bm \beta}:=\{ {\beta}_1, \dots , {\beta}_{N+1} \}. 
\label{QP}
\end{align} 
Their commutation relations, 
\begin{align}
[{\hat Q}({\bm \alpha}),{\hat P}_{\pm}({\bm \beta})]=i\left( \sum_{j=1}^{N}{\alpha}_j \beta_j
\mp {\alpha}_{N+1} {\beta}_{N+1} \right){\hat I},  
\label{CCCR}
\end{align}
lead to the weak (Heisenberg) form of the URs in product form:
\begin{align} 
\Delta Q({\bm \alpha})\, \Delta P_{\pm}({\bm \beta}) \geqq \frac{1}{2} 
\left| \sum_{j=1}^{N}{\alpha}_j {\beta}_j \mp {\alpha}_{N+1}{\beta}_{N+1} \right|.
\label{PUR}
\end{align}
This implies the sum-form inequalities
\begin{align}
\left[ \Delta Q({\bm \alpha}) \right]^2+\left[ \Delta P_{\pm}({\bm \beta}) \right]^2  \geqq 
\left| \sum_{j=1}^{N}{\alpha}_j {\beta}_j \mp {\alpha}_{N+1}{\beta}_{N+1} \right|.
\label{SUR} 
\end{align}
The variances of the EPR-like observables (\ref{QP}),  
\begin{align}
& \left[ \Delta Q({\bm \alpha}) \right]^2=\sum_{j=1}^{N+1} \sum_{k=1}^{N+1} {\alpha}_j {\alpha}_k 
{\sigma}(q_j, q_k)      \notag   \\
& -4{\alpha}_{N+1}\sum_{j=1}^{N} {\alpha}_j {\sigma}(q_j, q_{N+1}),      \notag  \\
& \left[ {\Delta P}_{+}({\bm \beta}) \right]^2
=\sum_{j=1}^{N+1} \sum_{k=1}^{N+1} {\beta}_j {\beta}_k {\sigma}(p_j, p_k),     \notag  \\
& \left[ {\Delta P}_{-}({\bm \beta}) \right]^2
=\sum_{j=1}^{N+1} \sum_{k=1}^{N+1} {\beta}_j {\beta}_k {\sigma}(p_j, p_k)    \notag  \\
& -4{\beta}_{N+1}\sum_{j=1}^{N} {\beta}_j {\sigma}(p_j, p_{N+1}) 
\label{varQP}
\end{align} 
are quadratic forms in the positive variables ${\bm \alpha}$ and ${\bm \beta}$, respectively.
Their coefficients are covariances of the canonical quadrature operators 
of the $(N\, \text{vs} \,1)$-mode state $\hat \rho$: 
\begin{align}
& \sigma (q_j, q_k):=\langle \hat q_j\hat q_k\rangle -\langle \hat q_j \rangle \langle \hat q_k \rangle,  
\notag   \\
& \sigma (p_j ,p_k):=\langle \hat p_j\hat p_k\rangle -\langle \hat p_j \rangle\langle \hat p_k \rangle,
\notag   \\
& (j,k=1,\dots, N+1).
\label{mom} 
\end{align}
These are the entries of the covariance matrix (CM) of the two-party state $\hat \rho$, 
which is denoted ${\mathcal V} \in M_{2(N+1)}({\mathbb R})$, is symmetric, 
and has the block structure:
\begin{align}
{\mathcal V}=\left(
\begin{matrix} 
{\mathcal V}_{A}\;   &   \; {\mathcal C} \\   \\
{\mathcal C}^{\rm T} \; &   \;  {\mathcal V}_{B} 
\end{matrix} \right).
\label{partCM}
\end{align}
The partition (\ref{partCM}) allows one to vizualize the contribution of each subsystem.
Thus, $\mathcal V_{A}$ is the $2N \times 2N$ CM of the reduced $N$-mode state 
held by Alice, $\mathcal V_{B}$ is the $2 \times 2$ CM of the reduced one-mode state 
handled by Bob, while ${\mathcal C}$ is the $2N \times 2$ matrix describing 
the cross-correlations between the parties. 

Notice that the covariances $\sigma(q_j, p_k)$ of any coordinate-momentum pair 
do not appear in the quadratic forms (2.5). Apparently, the EPR-like inequalities (2.4) hold 
only for ($N$ vs 1)-mode state having a standard-form CM with vanishing covariances 
$\sigma(q_j, p_k)$. One could ask whether this might be a restriction on the class of states 
specified at the beginning of this section. This is not the case, because it was shown 
in Refs. \cite{Duan,SA2007,GK2014} that any multimode CM can be transformed 
to the standard form described above with local symplectic matrices such as one-mode 
rotations and squeezings.  Since two multimode states which are similar via a tensor product 
of one-mode unitaries possess the same amount of entanglement or steerability, they are 
equivalent as nonlocal resources. Therefore, our discussion applies to all the multimode states 
whose CMs are connected by one-mode symplectic transformations. The particular form 
acquired by the CM (\ref{partCM}) discussed above is very convenient in many aspects. 
For instance, due to the identity $\sigma(q_j, p_k)=0,$ it is more productive for subsequent 
evaluations to write the CM ${\cal V}$ as the direct sum 
\begin{align}    
{\cal V}={\cal V}^{(q)} \oplus {\cal V}^{(p)}.
\label{Vq+Vp}      
\end{align}
The two terms of the above decomposition are the $(N+1) \times (N+1)$ CMs 
in the position and momentum spaces, written with the reordered canonical operators 
$\{ {\hat q}_1, {\hat q}_2,  \dots , {\hat q}_{N+1} \}$ 
and $\{ {\hat p}_1, {\hat p}_2, \dots , {\hat p}_{N+1} \}$, 
respectively. A direct consequence of Eq. (\ref{Vq+Vp}) is the factorization rule
\begin{equation}  
\det( {\cal V} )=\det\left( {\cal V}^{(q)} \right) \det\left( {\cal V}^{(p)} \right).
\label{detCM}      
\end{equation}

Let us recall the restrictions imposed by the canonical commutation relations to the CM 
${\mathcal V}$ of an arbitrary $n$-mode state \cite{Simon0,Simon}. Essentially, 
any {\em bona fide} CM ${\mathcal V}$  fulfills the Robertson-Schr\"odinger 
matrix uncertainty relation, namely, the positive semidefiniteness condition
\begin{align}
{\mathcal V}+\frac{i}{2} J \geq 0, 
\label{RS}
\end{align} 
where  $J$ is the standard matrix of the symplectic form on ${\mathbb R}^{2n}$: 
\begin{align}
J:=\bigoplus_{k=1}^{n}\,J_k, \;\;  J_k:=\left( 
\begin{array}{cc}
0  & 1\\ -1 & 0
\end{array}
\right), \;\;\;  (k=1,\dots, n).
\label{J}
\end{align}
We stress that the conditions (\ref{PUR}), (\ref{SUR}), and (\ref{RS}) stem from the URs for 
the canonical observables and hence they have to be satisfied by any physical multimode state. 

The quantumness requirement (\ref{RS}) implies, via Williamson's theorem \cite{Williamson}, 
the positive definiteness of the CM ${\mathcal V}$, 
\begin{align} 
{\mathcal V}>0, 
\label{V>0}
\end{align} 
as well as the inequality 
\begin{align}
\det({\mathcal V}) \geqq \frac{1}{2^{2n} },  \qquad  (n=1, 2, 3, \dots).
\label{Williamson}
\end{align}

It is worth mentioning that the inequality (\ref{Williamson}) can be recovered 
by selecting from the whole class of $n$-mode states possessing the same CM 
${\mathcal V}$ a Gaussian one, denoted ${\hat \rho}_G$, which is determined 
up to a translation in the phase space. Indeed, its purity depends only on the determinant 
of the CM ${\mathcal V}$, as follows \cite{PTH2001,PTH2003}: 
\begin{align} 
{\rm Tr} \left[ \left( {\hat \rho}_G \right)^2 \right]=\frac{1}{2^n \sqrt{\det({\cal V}) } } \leqq 1.
\label{purG} 
\end{align}
This universal property of the purity, when applied to any $n$-mode Gaussian state, 
Eq. (\ref{purG}), coincides therefore with the general inequality (\ref{Williamson}). 

\section{EPR-like separability bounds for \\ 
bipartite ${\bm (}{\bm N} \, \text{vs} \, {\bm 1}{\bm )}$-mode states}

Use of uncertainty relations between nonlocal operators of the type (\ref{lm}) appeared 
to be a fruitful idea in quest of separability conditions for continuous-variable two-party states 
and an enforcement of the role of uncertainty principle in understanding quantum correlations. 
Besides the original approach of Duan {\em et al.} \cite{Duan}, some other necessary 
conditions of separability which employ pairs of more general nonlocal observables 
depending on one or more parameters have been written \cite{Simon,GMVT}.  
Essentially, all these treatments rely on the important finding that for separable states
the uncertainty relations should modify to become stronger. This was first shown 
in Refs. \cite{Duan,Simon}  by directly exploiting the expansion of a two-party separable state 
shared by Alice and Bob as a convex combination of product states \cite{Werner}:
\begin{align}
\hat \rho_{sep}:=\sum_{k} w_k {\hat \rho}_A^{(k)}\otimes {\hat \rho}_B^{(k)}, \quad  
(w_k >0, \;\;  \sum_{k} w_k=1). 
\label{sep}
\end{align}
More generally,  by applying appropriate URs, necessary conditions for separability 
of various kinds of partitions in multimode states were written in terms of variances 
of linear combinations of canonical quadrature operators \cite{LF2003}.  Their violations 
are sufficient for genuine multipartite entanglement.  Note that the line of research proposed 
in Ref. \cite{LF2003} led to the understanding of Gaussian cluster states \cite{LWG2007}.

The enhancement of URs for separable states can be  efficiently displayed  by using 
an important  theorem proven by Hofmann and Takeuchi \cite{HT} regarding 
the LURs as a tool for entanglement detection on any bipartite quantum state.  
Let us denote by $\hat {\cal  A}_k, \; \hat {\cal  B}_k$ some observables at our choice 
acting on Alice's  and Bob's side, respectively. The incompatible observables 
of each party obey the sum-type LURs   
\begin{align}
\sum _k \left( \Delta {\cal  A}_k \right)^2 \geqq C_A,  \qquad 
\sum _k \left( \Delta  {\cal  B}_k \right)^2 \geqq C_B,
\label{LUR}   
\end{align}
where $C_A$ and $C_B$ denote their local positive uncertainty limits. The theorem 
of Hofmann and Takeuchi \cite{HT} asserts that, in any separable bipartite state, 
the total sum-form uncertainty of the nonlocal observables
\begin{align}
\hat  {\cal  M}_k:=\hat {\cal  A}_k \otimes {\hat I}_k+{\hat I}_k \otimes \hat {\cal  B}_k 
\label{nonlocal}   
\end{align}
satisfies the inequality
\begin{align}
\sum _k \left( \Delta  {\cal  M}_k \right)^2 \geqq C_A+C_B.
\label{HT}   
\end{align}
In other words, for any separable bipartite state, the total sum uncertainty of the nonlocal 
observables (\ref{nonlocal}) is  minorized by the sum of the local uncertainty limits of both parties. 
The bound specified by the LURs  in Eq.(\ref{HT}) is the expression of a strong separability 
condition \cite{G2006}. For discrete-variable systems, the sum UR (\ref{HT}) can generally 
be formulated in terms of CMs built with the incompatible constituents of appropriate 
nonlocal observables  $\hat{\cal  M}_k$ \cite{G1,G2,NHa1}. Violation of any sum UR 
of the form (\ref{HT}) is therefore a signature of entanglement.
 
In continuous-variable systems, we have the privilege to employ in this approach 
the one-mode canonical observables $\hat q_j$ and $\hat p_j$, whose non-commutativity 
generates the Heisenberg uncertainty relations that are at the heart of quantum mechanics 
since its early days. The system of $(N\, \text{vs} \,1)$ modes we are dealing with 
is a perfect test bed for an elegant application of the Hofmann-Takeuchi inequality (\ref{HT}). 

We now focus on a separable $(N\, \text{vs} \,1)$-mode state ${\hat \rho}_{sep}$, 
with a partitioned CM ${\mathcal V}$, Eq. (\ref{partCM}), and choose as nonlocal observables 
$\hat {\cal M}_k$ the pair of generalized Reid operators (\ref{QP}): 
\begin{align}
\hat{\cal M}_1={\hat Q}(\bm{\alpha}),  \qquad   \hat{\cal M}_2={\hat P}_{\pm}(\bm{\beta}),
\label{QP(Nvs1)}   
\end{align}
whose one-party components are respectively:  
\begin{align} 
& \hat {\cal  A}_1= \sum_{j=1}^{N} {\alpha}_j{\hat q}_j,    \qquad
\hat {\cal  B}_1=-{\alpha}_{N+1} {\hat q}_{N+1};      \notag   \\
& \hat {\cal  A}_2=  \sum_{j=1}^{N}{\beta}_j {\hat p}_j,     \qquad
\hat {\cal  B}_2= \pm {\beta}_{N+1} {\hat p}_{N+1},    \notag   \\
& ( {\alpha}_j >0, \;\;  {\beta}_j >0: \;  j=1, \dots , N+1).
\label{Nvs1} 
\end{align} 
 
When inserting into the inequality (\ref{HT}) the sum-type LURs (\ref{LUR}) associated 
to the commutators  
\begin{align}   
&\left[ \hat{\cal  A}_1, \hat{\cal  A}_2 \right]=i\, C_A {\hat I}_A,    \qquad 
C_A :=\sum_{j=1}^N {\alpha}_j {\beta}_j,     \notag  \\
& \left[ \hat{\cal B}_1, \hat{\cal B}_2 \right]=\mp i\, C_B {\hat I}_B,    \qquad 
C_B:={\alpha}_{N+1} {\beta}_{N+1},
\label{lcom}
\end{align}
we get a pair of sum-form necessary conditions of separability: 
\begin{align}
\left[ \Delta Q({\bm \alpha}) \right]^2 +\left[ \Delta P_{\pm}({\bm \beta}) \right]^2
\geqq  \sum_{l=1}^{N+1}{\alpha}_l{\beta}_l.
\label{sepSUR}
\end{align}
The upper Eq.(\ref{sepSUR}) has a larger separability bound than the upper physicality bound 
in the corresponding sum-form UR (\ref{SUR}). Interestingly, the lower separability condition
(\ref{sepSUR}) coincides with the lower sum-form UR (\ref{SUR}).
Therefore, in order to be separable, an $(N\, \text{vs} \,1)$-mode state ${\hat \rho}$ has to obey 
both conditions (\ref{sepSUR}) for any values of the involved parameters 
$\{ {\bm \alpha}, {\bm \beta} \}$.

The separability conditions (\ref{sepSUR}) are always met when the absolute minima 
of the functions
\begin{align} 
{\Sigma}_{\pm}({\mathcal V};\,{\bm \alpha}, {\bm \beta})
:=\frac{\left[ \Delta Q({\bm \alpha}) \right]^2 +\left[ \Delta P_{\pm}({\bm \beta}) \right]^2} 
{\sum_{l=1}^{N+1}{\alpha}_l{\beta}_l} 
\label{sepS}
\end{align}
with respect to all their variables ${\alpha}_j >0, \;  {\beta}_j >0, \; ( j=1,\dots , N+1)$
are greater than or at least equal to $1$. Let us denote the corresponding minimum points
$\{ \bm{\alpha}_{\pm}, \bm{\beta}_{\pm} \}$ in order to write concisely the EPR-like
necessary conditions of separability (\ref{sepSUR}):
\begin{align} 
{\Sigma}_{\pm}({\mathcal V};\,{\bm \alpha}_{\pm}, {\bm \beta}_{\pm}) \geqq 1. 
\label{sepnc}
\end{align} 
Except for the simplest case $N=1$ \cite{PT2018},
an analytic evaluation of such a minimum is hampered by algebraic difficulties.
Nevertheless, in view of the formulas (\ref{varQP}), we apply Euler's theorem 
on homogeneous functions to the fractions (\ref{sepS}) and find the identities:
\begin{align}  
& \sum_{j=1}^{N+1} {\alpha}_j \frac{\partial}{\partial{\alpha}_j }
\left[ {\Sigma}_{\pm}({\mathcal V};\,{\bm \alpha}, {\bm \beta}) \right]        
=-\sum_{j=1}^{N+1} {\beta}_j \frac{\partial}{\partial{\beta}_j }
\left[ {\Sigma}_{\pm}({\mathcal V};\,{\bm \alpha}, {\bm \beta})  \right]        \notag    \\
& =\frac{\left[ \Delta Q({\bm \alpha}) \right]^2 -\left[ \Delta P_{\pm}({\bm \beta}) \right]^2} 
{\sum_{l=1}^{N+1}{\alpha}_l{\beta}_l}. 
\label{Euler}
\end{align}
Accordingly, any $(N\, \text{vs} \,1)$-mode separable state ${\hat \rho}_{sep}$ fulfills 
the conditions
\begin{align} 
\left[ \Delta Q(\bm{\alpha}_{\pm}) \right]^2 =\left[ \Delta P_{\pm}(\bm{\beta}_{\pm}) \right]^2
\geqq \frac{1}{2}\sum_{l=1}^{N+1}( {\alpha}_l )_{\pm}( {\beta}_l )_{\pm}.
\label{min_pm}
\end{align}

The difference between the necessary conditions of separability (\ref{min_pm})
for an $(N\, \text{vs} \,1)$-mode state ${\hat \rho}$ arises from the opposite signs
of the ${\hat p}_{N+1}$ terms in the nonlocal observables ${\hat P}_{\pm}({\bm \beta})$  
introduced in Eq. (\ref{QP}). As shown by Simon \cite{ Simon}, the change of sign 
$p_{N+1} \to  -p_{N+1}$ in the Wigner function  
$W(q_1, \dots , q_{N+1}, p_1,  \dots , p_{N+1})$ of the $(N\, \text{vs} \,1)$-mode state 
${\hat \rho}$ amounts to its transformation into the Wigner function of the partially 
transposed operator ${\hat \rho}^{ {\rm PT}(B) }$ with respect to Bob's party:
$$W^{ {\rm PT}(B) }(q_1, \dots , q_N, q_{N+1}, p_1, \dots , p_N, p_{N+1})$$
$$=W(q_1, \dots , q_N, q_{N+1}, p_1, \dots , p_N, -p_{N+1}).$$
By virtue of the decomposition (\ref{Vq+Vp}), the CM  of the partially transposed operator 
${\hat \rho}^{ {\rm PT}(B) }$ is positive definite: ${\mathcal V}^{ {\rm PT}(B) }>0.$
Moreover, in view of Ref. \cite{LSA2018}, it fulfils the Robertson-Schr\"odinger matrix UR:
\begin{align}
{\mathcal V}^{ {\rm PT}(B) }+\frac{i}{2} J \geq 0.
\label{RSUR}
\end{align}
As a matter of fact, Eq. (\ref{RSUR}) is a consequence of the Peres theorem \cite{Peres}, 
which asserts that any bipartite separable state ${\hat \rho}$ is a PPT state: this simply means 
that its partial transpose is positive, ${\hat \rho}^{ {\rm PT}(B) } \geq 0$. Therefore, 
Eq. (\ref{min_pm}) displays the quantumness requirements for the couple 
of $(N\, \text{vs} \,1)$-mode states ${\hat \rho}$ and ${\hat \rho}^{ {\rm PT}(B) }$.

The two-mode states $(N=1)$ are important in their own right. They  are present in many 
areas, generating a productive research in quantifying bipartite quantum correlations 
especially in the Gaussian scenario \cite{ARL2014,PTH2001,PTH2003,PT2008,PT2016}.
In the two-mode case, the EPR-like observables (\ref{QP}) depend on two pairs of positive
parameters, 
$${\bm \alpha}:=\{ {\alpha}_1, \, {\alpha}_2 \}, \qquad  {\bm \beta}:=\{ {\beta}_1, \, {\beta}_2\},$$ 
as follows:
\begin{align}
{\hat Q}({\bm \alpha}):={\alpha}_1 {\hat q}_1- {\alpha}_2 {\hat q}_2,    \quad
{\hat P}_{\pm}({\bm \beta}):= {\beta}_1 {\hat p}_1 \pm {\beta}_2 {\hat p}_2.
\label{QP1}
\end{align} 
Clearly, Reid's pair of nonlocal observables (\ref{lm}) built with two independent parameters, 
$\lambda$ and $\mu$, as well as the single-parameter ones employed by Duan {\em et al.} 
in Ref.\cite{Duan} are particular forms of EPR-like observables (\ref{QP1}).   

Let us consider an arbitrary separable two-mode state ${\hat \rho}_{sep}$, whose $4 \times 4$ 
CM ${\mathcal V}$, Eq. (\ref{partCM}), is always characterized by four standard-form 
parameters \cite{Duan}: 
\begin{align}
& b_1:=\sigma (q_1, q_1)=\sigma (p_1, p_1), \;\;\;  b_2:=\sigma (q_2, q_2)=\sigma (p_2, p_2), 
\notag    \\
& c:=\sigma (q_1, q_2), \;\;\;  d:=\sigma (p_1, p_2).
\label{bbcd}
\end{align}
With no loss of generality, one can choose: 
$$b_1 \geqq b_2 \geqq \frac{1}{2}, \qquad  c \geqq |d|.$$
Then the $2 \times 2$  submatrices in the partition (2.7) of the CM 
${\mathcal V}$ are diagonal:
\begin{align}
{\mathcal V}_A=b_1 I_2,   \qquad   {\mathcal V}_{B}=b_2 I_2,   \qquad  
{\mathcal C}=\left(
\begin{matrix} 
c\;   &   \;   0 \\   \\
0 \;  &   \;   d 
\end{matrix} \right), 
\label{sfCM}
\end{align}
where $I_2$ is the $2 \times 2$ identity matrix. We specialize Eq. (\ref{varQP}) to get
the variances of the EPR-like observables (\ref{QP1}):
\begin{align}
& \left[ \Delta Q({\bm \alpha}) \right]^2=b_1{\alpha}_1^2  +b_2{\alpha}_2^2  
-2c\, {\alpha}_1{\alpha}_2,      \notag   \\
& \left[ {\Delta P}_{\pm}({\bm \beta}) \right]^2=b_1{\beta}_1^2  +b_2{\beta}_2^2 
\pm 2d\, {\beta}_1 {\beta}_2.
\label{varQP1}
\end{align} 
The next step is to evaluate the absolute minima of the functions (\ref{sepS})
written for two-mode states,
\begin{align} 
{\Sigma}_{\pm}({\mathcal V};\,{\bm \alpha}, {\bm \beta})
:=\frac{\left[ \Delta Q({\bm \alpha}) \right]^2 +\left[ \Delta P_{\pm}({\bm \beta}) \right]^2} 
{ {\alpha}_1{\beta}_1+{\alpha}_2 {\beta}_2 }, 
\label{sepS1}
\end{align}
with respect to their four positive variables $\{ {\bm \alpha}, {\bm \beta} \}.$ 
These minima are found to be equal to the doubles of the smallest symplectic 
eigenvalues ${\kappa}_{-}^{\rm PT}$ and ${\kappa}_{-}$ \cite{VW2002} 
of the CMs ${\mathcal V}^{\rm PT}$ and ${\mathcal V}$, respectively:
\begin{align} 
{\Sigma}_{+}({\mathcal V};\,{\bm \alpha}_{+}, {\bm \beta}_{+})=2{\kappa}_{-}^{ {\rm PT} },   
\quad   {\Sigma}_{-}({\mathcal V};\,{\bm \alpha}_{-}, {\bm \beta}_{-})=2{\kappa}_{-}.
\label{minS1}
\end{align}
In order to derive the formulas (\ref{minS1}) one needs the symplectic eigenvalues 
${\kappa}_{\pm}$ and ${\kappa}_{\pm}^{\rm PT}$ expressed in terms 
of the standard-form parameters (\ref{bbcd}) \cite{PT2018a}:
\begin{align} 
({\kappa}_{\pm})^2=\frac{1}{2}\left( b_1^2 + b_2^2 +2cd \pm \sqrt{\Delta} \right),
\label{kappa}
\end{align}
where $\Delta$ is the discriminant of a quadratic trinomial:
\begin{align} 
& \Delta=\left( b_1^2 +b_2^2 +2cd \right)^2 -4\det({\mathcal V})    \notag   \\
& =\left( b_1^2 -b_2^2 \right)^2 +4(b_1 c+b_2 d)(b_2 c+b_1 d) \geqq 0.
\label{Delta}
\end{align}
Similarly, 
\begin{align} 
\left( {\kappa}_{\pm}^{\rm PT} \right)^2=\frac{1}{2}\left( b_1^2 + b_2^2 
-2cd \pm \sqrt{ {\Delta}^{\rm PT} }\right),
\label{kappaPT}
\end{align}
with the discriminant
\begin{align} 
& {\Delta}^{\rm PT} =\left( b_1^2 +b_2^2 -2cd \right)^2 -4\det({\mathcal V})    \notag   \\
& =\left( b_1^2 -b_2^2 \right)^2 +4(b_1 c-b_2 d)(b_2 c-b_1 d) \geqq 0.
\label{DeltaPT}
\end{align}
Although recalled in Ref. \cite{PT2018} just in the special case of separable 
two-mode Gaussian states, Eqs. (\ref{kappa})-(\ref{DeltaPT}) hold for any separable 
two-mode state, either Gaussian or non-Gaussian. It is worth mentioning
that they imply the following sign rule:
\begin{align} 
\sgn\left( {\kappa}_{-}^{\rm PT} -{\kappa}_{-} \right)=\sgn(d).
\label{sgn(d)}
\end{align}

According to Eq. (\ref{minS1}), the EPR-like necessary conditions of separability 
(\ref{sepnc}) acquire in the two-mode case their simplest form:
\begin{align} 
{\kappa}_{-}^{\rm PT}  \geqq \frac{1}{2},  \qquad  {\kappa}_{-}  \geqq \frac{1}{2}.
\label{sepnc1}
\end{align}
The first above inequality manifestly illustrates Peres' criterion of inseparability \cite{Peres},
while the second one is just the quantumness requirement for the given separable 
two-mode state ${\hat \rho}_{sep}$. Note that the first minimal normalized UR sum 
in Eq. (\ref{minS1}) was derived in Ref. \cite{PT2018} by using a special set 
of independent variables, in the framework of a comprehensive analysis.
 
\section{Uncertainty-relation criterion for EPR steering}

Our aim in this section is to establish the sum-form URs valid for {\em unsteerable} 
two-party states. To this end, we briefly recall the meaning of the steering scenario, as emerging  
from its interpretation in Refs.\cite{W2,W3,Pusey,SNC,QVB,UMG,PW,Adesso2015,Kiukas}.
Accordingly, two separate observers, Alice and Bob, share a  bipartite state ${\hat \rho}_{AB}$ 
of a composite quantum system, such that its Hilbert space is the tensor product 
${\cal H}_A \otimes {\cal H}_B$. Alice performs a local unknown measurement $x$ whose 
positive  operators $\hat M_{a|x}$ add up to identity:  $\sum_{a} \hat M_{a|x}={\hat I}_A$. 
Here $a$ denotes a possible outcome of the measurement chosen by Alice on her side, 
having the probability $p(a|x)$. The conditional unnormalized state on Bob's side, 
\begin{align}  
{\hat \sigma}^B(a|x)
={\rm Tr}_A \left[ {\hat \rho}_{AB} \left( {\hat M}_{a|x} \otimes {\hat I}_B \right) \right],
\label{as}
\end{align}
has the trace ${\rm Tr}_B \left[ {\hat \sigma}^B(a|x) \right] =p(a|x)$. 
Bob's collection of states $\{{\hat \sigma}^B(a|x)\}_{a, x}$ built with all the possible pairs $\{a,x \}$ 
is called an {\em assemblage} \cite{Pusey}. Let us write the sum rule
\begin{align}  
\sum_a {\hat \sigma}^B(a|x) ={\rm Tr}_A \left( {\hat \rho}_{AB} \right)=:{\hat \rho}_B,
\label{sumas}
\end{align}
giving the reduced state at Bob's hand, prior to Alice's measurement. Equation (4.2) shows 
that Bob's reduced state ${\hat \rho}_B$ is not modified whatever the {\em nonselective}
measurement $x$ made by Alice.

When all the conditional states (\ref{as}) have the special structure
\begin{align} 
{\hat \sigma}^B_{\rm us}(a|x)=\sum_{\lambda} p(a|x, \lambda) q(\lambda) {\hat \rho}_B(\lambda),
\label{as1}
\end{align}
then the assemblage $\{ {\hat \sigma}^B_{\rm us}(a|x) \}_{a, x}$ is termed {\em unsteerable}. 
In Eq.(4.3),  $\{ {\hat \rho}_B(\lambda) \}_{\lambda}$ is a preexisting ensemble of local hidden 
states (LHSs), each one depending on the local hidden variable (LHV) $\lambda$ 
and having the associated weight $q(\lambda)$. The weights are normalized, {\em i.e}, 
they sum to one: $\sum_{\lambda}q(\lambda)=1$. Further, $p(a|x, \lambda)$ is the probability of 
the outcome $a$ in Alice's measurement $x$, provided that Bob's party is described by the LHS
$\{ {\hat \rho}_B(\lambda) \}$. The trace of the operator (4.3) acting on the Hilbert space 
${\cal H}_B$ is just the probability of the outcome $a$ in Alice's measurement $x$:
\begin{align} 
p(a|x)= \sum_{\lambda} p(a|x, \lambda) q(\lambda).
\label{tras1}
\end{align}
On the other hand,  the initial  Bob's reduced state has the convex decomposition
\begin{align}  
{\hat \rho}_B=\sum_{\lambda} q(\lambda) {\hat \rho}_B(\lambda).
\label{rhoB}
\end{align}  
When the local measurement $x$ performed by Alice has the outcome $a$, then it changes 
Bob's reduced state (\ref{rhoB}) into an updated convex combination of LHSs:
\begin{align}  
{\hat \rho}_B (a|x)=\frac{1}{p(a|x)}{\hat \sigma}^B_{\rm us}(a|x)
=\sum_{\lambda} \frac{p(a|x,\lambda) q(\lambda)}{p(a|x)} {\hat \rho}_B(\lambda) .
\label{rhoBax}
\end{align} 

The key quantity in the analysis of unsteerability is the joint probability of a pair of outcomes 
$\{ a,\, b \}$ in two successive measurements, $x$, performed by Alice, and $y$, made by Bob:
\begin{align}  
P\left( a|x, \, b|y, \,  {\hat \rho}_{AB} \right)
={\rm Tr}_B\left[ {\hat \sigma}^B(a|x) {\hat M}_{b|y}\right].
\label{P(ab)}
\end{align}
In view of Eq. (4.3), for any unsteerable assemblage
$\left\{{\hat \sigma}^B_{\rm us}(a|x) \right\}_{a, x}$, Eq. (4.7) acquires 
a specific form \cite{W1,W2}:
\begin{align} 
& P\left( a|x, \, b|y, \,  {\hat \rho}_{AB} \right)=\sum_{\lambda} q(\lambda) p(a|x, \lambda) 
P\left[ b|y, \, {\hat \rho}_B(\lambda) \right].
\label{P(ab)us}
\end{align}    
Thus, the joint probability (\ref{P(ab)}) has the asymmetric structure (\ref{P(ab)us}) 
whenever the measurement performed by Alice cannot steer (pilot) Bob's party. Essentially,
Eq. (\ref{P(ab)us}) can be interpreted in terms of a LHS model for Bob and a LHV model
for Alice, which is correlated with Bob's reduced state (\ref{rhoB}) via the conditional 
probabilities $p(a|x, \lambda)$. The lack of such a model even for a single pair $\{a,x\}$ 
of the assemblage means steering from Alice to Bob  \cite{W1,W2,W3}.

It can be directly seen that any measurement  performed by Alice on a separable state, 
Eq. (\ref{sep}), leads to an unsteerable assemblage (\ref{as1}). It follows that 
entanglement is a necessary condition for EPR steerability. For instance, an unsteerable 
assemblage can be obtained from the separable state  ${\hat \rho}_{AB}
=\sum_{\lambda} q(\lambda)\left( |\lambda \rangle \langle \lambda| \right)_A 
\otimes {\hat \rho}_B(\lambda)$. Here, Alice performs projective measurements 
of mutually commuting observables whose common eigenvectors are generically denoted 
$|\lambda \rangle$. Moreover, it was proven that unsteerable assemblages can be obtained 
by joint local measurements \cite{QVB,UMG,Kiukas}. Joint measurability is an extension 
of commutativity to general measurements described by positive operator valued measures 
(POVMs). Note that observation of one-way EPR steering requires the presence of both 
entanglement and incompatible measurements. The connection between steering 
and incompatibility was recently extended to continuous-variable systems \cite{Kiukas,Kiukas2}. 
To conclude this discussion, an unsteerable assemblage of the type (\ref{as1}) at Bob's party 
is provided by Alice via local measurements of commuting or jointly measurable observables. 

The main question we have now to answer is as follows: how might URs be written analogously
to the Hofmann-Takeuchi inequality (\ref{HT}) such as to be specifically valid for two-party 
unsteerable states? The above-sketched tableau of one-way EPR steering offers an idea 
for answering this question. First, let us remark that the structure (\ref{as1}) of an unsteerable 
assemblage from Alice to Bob imposes the existence of a local uncertainty limit $C_B$. 
Second, this is not the case for Alice's side, where no uncertainty relation is available 
since her measurements are unknown. In other words, our lack of knowledge about 
the nature of Alice's measurements amounts to ignoring the uncertainty limit $C_A$ 
of her party \cite{NHa1,Zhen,Zhen1}. We therefore suggest that the right necessary 
condition for unsteerabilty of Bob's reduced state through Alice's local measurement i
s the sum-form UR  
\begin{align}
\sum _k \left( \Delta {\cal  M}_k \right)^2 \geqq C_B,
\label{ABus}
\end{align}
which is  weaker than the similar one for separability, Eq. (\ref{HT}). Any violation of the inequality
(\ref{ABus}) is thus a pertinent signature of steering from Alice to Bob. 

Accordingly, with the EPR-like observables (\ref{QP}), one gets the sum-form UR which is compulsory
for any $(N\, \text{vs} \,1)$-mode state which is unsteerable ($us$) from Alice to Bob:
\begin{align}   
\left[ \Delta Q({\bm \alpha}) \right]^2+\left[ \Delta P_{+}({\bm \beta}) \right]^2
\geqq {\alpha}_{N+1}{\beta}_{N+1}.
\label{usAB}
\end{align}
Similarly, for the same state, the necessary condition of unsteerability from Bob to Alice reads:       
\begin{align} 
\left[ \Delta Q(\bm{\alpha}) \right]^2+\left[ \Delta P_{+}(\bm{\beta}) \right]^2 
\geqq {\sum_j^{N}\alpha_j \beta_j}. 
\label{usBA}
\end{align}
Conditions (\ref{usAB}) and (\ref{usBA}) display the asymmetry of steering. 
They have to be fulfilled for all the positive parameters $\{ {\bm \alpha}, {\bm \beta} \}$. 
If the conditions (\ref{usAB}) or (\ref{usBA}) are not met for some values of these parameters, 
then the $(N \, \text{vs} \,1)$-mode state is steerable from Alice to Bob or from Bob to Alice, 
respectively. 
 
In the next section we intend to find the extremal normalized URs derived from
Eqs. (\ref{usAB}) and (\ref{usBA}).  Analytically, these distinct extremization problems 
for the two ways of steering  depend on $2N+2$ parameters and appear to be 
less complicated than those of the functions (\ref{sepS}) for separability.

\section{EPR-like uncertainty limits for steering}

We carry out here the extremization of the two properly normalized uncertainty sums 
required for the unsteerability of a $(N\, \text{vs} \,1)$-mode state in both possible ways 
of steering.

\subsection{Steering from Alice to Bob}
 
Guided by the UR (\ref{usAB}) we define the normalized sum
\begin{align} 
{ \Sigma}^{(A \to B)}_{\rm us}({\bm \alpha}, {\bm \beta})
:=\frac{\left[ \Delta Q({\bm \alpha}) \right]^2
+\left[ \Delta P_{+}({\bm \beta}) \right]^2}{ {\alpha}_{N+1}{\beta}_{N+1} } \geqq 1. 
\label{normS}
\end{align}
On account of the expressions (\ref{varQP}) of the variances involved in the above function,
we first write its extremization conditions with respect to the variables $\alpha_j,\beta_j$:
$$\frac{\partial { \Sigma}^{(A \to B)}_{\rm us}}{\partial \alpha_j}=0, \;\;\;
\frac{\partial { \Sigma}^{(A \to B)}_{\rm us}}{\partial \beta_j}=0, \;\;\; (j=1, \dots , N). $$
They amount to a couple of independent systems of $N$ linear equations:
\begin{align} 
\sum_{k=1}^N \alpha_k \sigma (q_j,q_k)- \alpha_{N+1}\sigma (q_j, q_{N+1})=0, 
\label{cond1ab}
\end{align}
and
\begin{align} 
\sum_{k=1}^N \beta_k \sigma (p_j,p_k)+ \beta_{N+1}\sigma (p_j, p_{N+1})=0.
\label{cond2ab}
\end{align}
The system (\ref{cond1ab}) with the variables $\alpha_j, \;\;  (j=1, \dots , N)$  has        
the coefficient matrix ${\cal V}_A^{(q)}$, while the system (\ref{cond2ab}) with the variables 
$\beta_j, \;\; (j=1, \dots , N)$ has the coefficient matrix ${\cal V}_A^{(p)}$. 

We start to minimize  the function (\ref{normS}) by using the conditions (\ref{cond1ab})
and (\ref{cond2ab}):
\begin{align} 
& {\Sigma}^{(A \to B)}_{\rm us}=\frac{1}{ {\beta}_{N+1} } \left[ {\alpha}_{N+1}\sigma (q_{N+1}, q_{N+1} )
-\sum_{j=1}^N {\alpha}_j \sigma (q_j, q_{N+1} ) \right]   \notag \\
& +\frac{1}{ {\alpha}_{N+1} } \left[ {\beta}_{N+1}\sigma (p_{N+1}, p_{N+1} )  
+\sum_{j=1}^N {\beta}_j \sigma (p_j, p_{N+1} ) \right].  \notag \\
\label{sum1}
\end{align}
When inserting into Eq.(\ref{sum1}) the solutions $\alpha_j$ and $\beta_j$ 
of the linear systems (\ref{cond1ab}) and (\ref{cond2ab}), written with Cramer's rule, 
we get two fractions whose numerators are proportional to the Laplace expansions 
of $\det( {\cal V}^{(q)} )$ and $\det( {\cal V}^{(p)} )$ along the $(N+1)$-th columns 
of their matrices. We find thus the function
 \begin{align} 
{ \Sigma}^{(A \to B)}_{\rm us}(\epsilon)=\epsilon\frac{\det{({\cal V}^{(q)})}}
{\det ({\cal V}_A^{(q)})}+\frac{1}{\epsilon} \frac{\det{({\cal V}^{(p)})}}{\det ({\cal V}_A^{(p)})},
\label{sum3}
\end{align}
of a single positive variable, $\epsilon:={\alpha}_{N+1}/{\beta}_{N+1}$.
Its unique minimum point,
$$\epsilon_m =\left[ \frac{\det{({\cal V}^{(p)})}} {\det{({\cal V}^{(q)})}}\frac{\det ({\cal V}_A^{(q)} )}                  
{\det ({\cal V}_A^{(p)})} \right]^{\frac{1}{2} },$$
gives, via Eq.(\ref{detCM}), the minimal uncertainty sum
\begin{align}
\min_{\{\bm{\alpha}, \bm{\beta} \} } { \Sigma}^{(A\rightarrow B)}_{\rm us}
= 2\, \sqrt{\frac{\det( {\mathcal V} ) }{\det( {\mathcal V}_A) } }.
\label{sm}
\end{align}
In view of the URs (\ref{usAB})  satisfied by unsteerable states from Alice to Bob we get 
the following necessary condition of unsteerability: 
\begin{align}
\frac{\det( {\mathcal V} ) }{\det( {\mathcal V}_A) } \geqq \frac{1}{4}.
\label{ncusAB}
\end{align}
Therefore, in order to be unsteerable from Alice, who operates on an $N$-mode state,  
to Bob, who holds a single-mode state, an $(N \, \text{vs} \,1)$-mode state has to obey 
the condition (\ref{ncusAB}). Its violation is a signature of steerability of {\em any} 
$(N \, \text{vs} \,1)$-mode state and an expression of the EPR-paradox as stated by Reid 
in Ref.\cite{Reid1}. 

\subsection{Steering from Bob to Alice}

For the same setting of $(N \, \text{vs} \,1)$-mode  we exploit the alternative way 
of unsteerability (\ref{usBA}) by using the condition imposed to the normalized sum-form UR 
\begin{align} 
{ \Sigma}^{(B \to A)}_{\rm us}(\bm{\alpha},\bm{\beta})
:=\frac{ \left[ \Delta Q({\bm \alpha}) \right]^2
+\left[ \Delta P_{+}({\bm \beta}) \right]^2}{\sum_{j=1}^{N} \alpha_j \beta_j} \geqq 1.
\label{normSb}
\end{align} 
It is convenient to start from its extremization conditions with respect to the variables 
$\alpha_{N+1}$ and $\beta_{N+1}$:
$$\frac{\partial { \Sigma}^{(B \to A)}_{\rm us}}{\partial \alpha_{N+1}}=0,  \qquad
 \frac{\partial { \Sigma}^{(B \to A)}_{\rm us}}{\partial \beta_{N+1}}=0,$$ 
which read, respectively:
\begin{align}
&  \alpha_{N+1}{\sigma (q_{N+1},q_{N+1})}={\sum_{k=1}^N \alpha_k \sigma (q_k, q_{N+1})}  
\label{cond1ba}
\end{align}
and 
\begin{align}
& \beta_{N+1}{\sigma (p_{N+1},p_{N+1})}=-\sum_{k=1}^N \beta_k \sigma (p_k, p_{N+1}).      
\label{cond2ba}
\end{align} 
Taking account of Eqs. (\ref{cond1ba}) and (\ref{cond2ba}), the variances (\ref{varQP}) 
acquire reduced forms that are characteristic for an $N$-mode state:
\begin{align}
& \left[ {\Delta Q}({\bm \alpha}^{\prime}) \right]^2:=\sum_{j=1}^{N}\sum_{k=1}^{N} 
\alpha_j \alpha_k \sigma^{\prime} (q_j,q_k),      \notag  \\
& \left[ {\Delta P}({\bm \beta}^{\prime}) \right]^2:=\sum_{j=1}^{N} \sum_{k=1}^{N} 
\beta_j \beta_k \sigma^{\prime} (p_j,p_k),          \notag  \\
& \text{with} \;\;\;  {\bm \alpha}^{\prime}:=\{ {\alpha}_1, \dots , {\alpha}_N \}, 
\quad  {\bm \beta}^{\prime}:=\{ {\beta}_1, \dots , {\beta}_N \}.
\label{newDelta}
\end{align} 
Here we have denoted: 
\begin{align} 
& \sigma^{\prime} (q_j,q_k):= \sigma (q_j,q_k)            \notag  \\
& -\sigma (q_j,q_{N+1}) [{\sigma (q_{N+1},q_{N+1})}]^{-1}\sigma (q_{N+1},q_k),     \notag  \\
& \sigma^{\prime} (p_j,p_k):=\sigma (p_j, p_k)            \notag  \\
& -\sigma (p_j,p_{N+1}) [{\sigma (p_{N+1},p_{N+1})}]^{-1}\sigma (p_{N+1},p_k),    \notag  \\ 
& (j, k=1, \dots  ,N).
\label{newCM} 
\end{align} 
Therefore, the unsteerability condition (\ref{normSb})  takes a simpler form: 
\begin{align}
& \min_{\{ {\bm \alpha}, {\bm \beta} \} } { \Sigma}^{(B\rightarrow A)}_{\rm us}
=\min_{\{ {\bm \alpha}^{\prime}, {\bm \beta}^{\prime} \}}
\frac{\left[ \Delta Q({\bm \alpha}^{\prime}) \right]^2+ \left[ \Delta P({\bm \beta}^{\prime}) \right]^2}
{\sum_{j=1}^{N} {\alpha}_j {\beta}_j} \geqq 1.    
\label{normSc}
\end{align} 
To see its significance, let us introduce two nonlocal operators built with the same positive 
parameters ${\bm \alpha}^{\prime}$ and ${\bm \beta}^{\prime}$ as in Eq. (\ref{newDelta}):
\begin{align} 
&{\hat Q}({\bm \alpha}^{\prime}):=\sum_{j=1}^{N}{\alpha}_j \hat q_j, \qquad
{\hat P}({\bm \beta}^{\prime}):=\sum_{j=1}^{N}{\beta}_j \hat p_j.
\label{genQPnew}
\end{align} 
Equations (\ref{newDelta}) appear to specify  the variances of the observables 
(\ref{genQPnew}) for the class of $N$-mode states which share the CM 
with the only non-vanishing entries (\ref{newCM}). 

Indeed, this matrix is precisely the Schur complement of the  $2\times 2$ CM ${\mathcal V_B}$ 
of the one-mode state held by Bob according to the partition (\ref{partCM}): 
\begin{align} 
{\mathcal V}/{\mathcal V_B}={\mathcal V}_A-{\mathcal C}\left( {\mathcal V_B} \right)^{-1}
{\mathcal C}^{\rm T}.
\label{Schur}
\end{align}    
Recall \cite{LSA2018} that the Schur complement (\ref{Schur}) satisfies the physicality condition 
\begin{align}    
{\mathcal V}/{\mathcal V_B}+\frac{i}{2}J_A \geq 0,
\label{BAusV/VB}
\end{align}
which qualifies it as a {\em bona fide} CM of an $N$-mode state \cite{Simon,Simon0}.  

By the same token, the inequality (\ref{normSc}) holds for the  above-mentioned class 
of $N$-mode states. In other words, Eqs. (\ref{normSc}) and (\ref{BAusV/VB}) 
are two {\em equivalent} necessary conditions of unsteerability from Bob to Alice.
They are simultaneously obtained by taking the minimum of the normalized 
sum-form UR (\ref{normSb}). 

Furthermore, we employ the Aitken factorization formula \cite{HJ} 
for the partitioned CM (\ref{partCM}):
\begin{align}
{\mathcal V}={\mathcal T}\, {\mathcal D}\, {\mathcal T}^{\rm T}.
\label{Aitken}
\end{align}
The matrix product (\ref{Aitken}) consists of a central factor, namely, the positive definite 
block-diagonal matrix,
\begin{align}
{\mathcal D}=\left( {\mathcal V}/{\mathcal V_B} \right) \oplus {\mathcal V}_{B},
\label{D}
\end{align}
sandwiched between a pair of unimodular triangular matrices,
\begin{align}
{\mathcal T}=\left(
\begin{matrix} 
I_A \;  &  \; {\mathcal C}{\mathcal V}_{B}^{-1} \\   \\
0 \;  &  \;  I_B   
\end{matrix} \right),   \quad
{\mathcal T}^{\rm T}=\left(
\begin{matrix} 
I_A \;  &  \;  0  \\   \\
{\mathcal V}_{B}^{-1}{\mathcal C}^{\rm T} \;  &   \;   I_{B} 
\end{matrix} \right).
\label{T}
\end{align}
The Aitken formula (\ref{Aitken}) is therefore a LDU decomposition, exhibiting 
the $^{\rm T}$congruence of the CM (\ref{partCM}) to the direct sum ${\mathcal D}$,
Eq. (\ref{D}), via the upper triangular matrix ${\mathcal T}$, Eq. (\ref{T}). The above discussion 
is valid for any bipartite $(N \, \text{vs} \,M)$ CM.  A straightforward consequence 
of Eqs. (\ref{Aitken})-(\ref{T}) is the Schur determinantal formula: 
\begin{align}  
\det ({\mathcal V}/{\mathcal V_B})=\frac{\det( {\mathcal V} )}{\det\left( {\mathcal V}_B \right)}.
\label{detSchur}  
\end{align}

Equations (\ref{Aitken})-(\ref{T}) clearly imply the equivalence between the matrix inequality  
(\ref{BAusV/VB}) and the following one:
\begin{align}    
{\mathcal V}+\frac{i}{2}J_A \oplus 0_B \geq 0.
\label{BAusV}
\end{align}
To sum up, we have proven that two necessary conditions of unsteerability from Bob to Alice,
Eqs. (\ref{normSc}) and (\ref{BAusV}), are fully equivalent, because each of them is equivalent
to Eq. (\ref{BAusV/VB}). This is the main result of the current subsection.

Owing to Williamson's theorem \cite{Williamson}, the physicality condition (\ref{BAusV/VB}) 
implies the inequality
\begin{align} 
\det \left( {\mathcal V}/{\mathcal V_B} \right) \geqq \frac{1}{2^{2N} }.
\label{detSCVB}
\end{align}
By combining Eq. (\ref{detSCVB}) with the Schur determinantal formula (\ref{detSchur}),
one gets a numerical necessary condition of unsteerability from Bob to Alice:
\begin{align}  
\frac{\det ({\mathcal V} )}{\det\left( {\mathcal V}_B \right)} \geqq \frac{1}{2^{2N} }.
\label{ncusBA} 
\end{align}
Except for the two-mode states $(N=1)$, this is weaker than the matrix condition (\ref{BAusV}).

The inequality (\ref{detSCVB}) is analog to the necessary condition of unsteerability 
from Alice to Bob (\ref{ncusAB}), written with Schur's formula (\ref{detSchur}) as  
\begin{align} 
\det \left( {\mathcal V}/{\mathcal V_A} \right) \geqq \frac{1}{4},
\label{detSCVA}
\end{align}
Note, however, that the numerical inequality (\ref{detSCVA}) is the only one
which is equivalent to the matrix necessary conditions of unsteerability from Alice to Bob 
similar to the equivalent matrix inequalities (\ref{BAusV/VB}) and (\ref{BAusV}):
\begin{align}    
{\mathcal V}/{\mathcal V_A}+\frac{i}{2}J_B \geq 0 \;  \iff   \;
{\mathcal V}+ 0_A \oplus \frac{i}{2}J_B \geq 0.
\label{ABusV}
\end{align}
The above extremizations of the normalized sums of EPR uncertainties 
do not allow us to find explicitly the corresponding minima. Nevertheless, violation 
of at least one of the similar conditions (\ref{ncusAB}) and (\ref{ncusBA}) is sufficient 
for the one-way steerability of {\em any} $(N \, \text{vs} \,1)$-mode state. Such violations 
are therefore signatures of steerability from Alice to Bob and, respectively, from Bob to Alice.

\section{Separability and unsteerability criteria for
${\bm (}{\bm N} \, \text{vs} \, {\bm 1}{\bm )}$-mode \\ Gaussian states revisited}

The class of Gaussian states (GSs) is a very special one in the whole set 
of continuous-variable multimode states. Any $n$-mode GS ${\hat \rho}_G$ 
is completely determined by a displacement vector in the phase space, 
$u \in {\mathbb R}^{2n}$, and the CM ${\mathcal V} \in M_{2n}({\mathbb R})$, 
which is subjected to the quantumness requirement (\ref{RS}). This characteristic
property of the GSs is the main reason for the central role they play in quantum
optics, as well as in quantum information theory.

\subsection{Separability criteria for Gaussian states}

Let us start with the case of the two-mode GSs. As recalled in Sec. III, 
the first EPR-type inequality in Eq. (\ref{sepnc}),
\begin{align} 
{\kappa}_{-}^{\rm PT} \geqq \frac{1}{2}, 
\label{PPTSimon}
\end{align}
is manifestly equivalent to the Peres-Simon PPT necessary condition of separability 
\cite{Peres,Simon}. Moreover, in Ref. \cite{Simon}, Simon proved that 
the physicality requirement (\ref{PPTSimon}) for the existence of the two-mode GS 
${\hat \rho}_G^{\rm PT}$ is also a criterion (sufficient condition) for the separability 
of a given two-mode GS ${\hat \rho}_G$. Accordingly, the symplectic eigenvalue 
${\kappa}_{-}^{\rm PT}$ is a PPT-type separability indicator at hand for any two-mode GS. 
Furthermore, its equivalence with a less convenient EPR-type separability indicator 
was explicitly proven \cite{PT2018b}.

The next accomplishment is the treatment of the separability problem for 
$(N \, \text{vs} \,1)$-mode GSs by Werner and Wolf in Ref. \cite{WW2001}.
Their main result is that the PPT property of such a GS implies its separability. 
Although this important Werner-Wolf theorem generalizes to an arbitrary $N$ 
the remarkable Simon's one for $N=1$ \cite{Simon}, it does not provide 
any analytic separability indicator for $N>1$. A second complementary result obtained 
in Ref. \cite{WW2001} is that the PPT property and separability are not equivalent
concepts for {\em any} $(N \, \text{vs} \, M)$-mode GS with $M>1$. Indeed, 
the authors get an example of a  $(2 \, \text{vs} \, 2)$-mode GS which is PPT, 
but entangled. Such a bipartite state is termed {\em bound entangled} \cite{H4}.
A comprehensive analysis of both above-mentioned results is recently developed 
in Ref. \cite{LSA2018}.

Coming back to the present work, we stress that, by virtue of the Werner-Wolf theorem,
the EPR-like necessary conditions of separability (\ref{min_pm}) are also {\em sufficient} 
ones. They could therefore be employed to get computable separability indicators 
for $(N \, \text{vs} \,1)$-mode GSs.

\subsection{Unsteerability criteria for Gaussian states}  

We recall the noteworthy results concerning the special case of Gaussian unsteerability
established by Jones, Wiseman, and Doherty in Ref. \cite{W2}.  Let us consider a bipartite 
$(N \,\text{vs} \, M)$-mode GS ${\hat \rho}_G$ shared by Alice and Bob, whose CM ${\mathcal V}$ 
has an adequate partition of the type (\ref{partCM}). Based on the assumption that Alice makes 
only Gaussian measurements, the necessary and sufficient condition for the Alice-to-Bob 
unsteerability of the GS ${\hat \rho}_G$ found in Refs. \cite{W1,W2} is
\begin{align}
{\mathcal V}+0_A \oplus  \frac{i}{2}J_B \geq 0. 
\label{JWD-AB}
\end{align}
Obviously, a similar necessary and sufficient condition holds for the Bob-to-Alice unsteerability 
of the GS ${\hat \rho}_G$:
\begin{align}
{\mathcal V}+ \frac{i}{2}J_A \oplus  0_B \geq 0. 
\label{JWD-BA}
\end{align}
As shown in Refs. \cite{Adesso2,Nha,LSA2018}, the unsteerability inequality (\ref{JWD-AB}) 
is equivalent to the physicality requirement
\begin{align}
{\mathcal V}/{\mathcal V_A}+\frac{i}{2} J_B \geq 0
\label{LSA}
\end{align}
for the Schur complement of the CM ${\mathcal V}_A$ in a partition of the type (\ref{partCM}),
\begin{align}
{\mathcal V}/{\mathcal V_A}={\mathcal V}_B-{\mathcal C}^{\rm T}
{\left( \mathcal V_A \right)}^{-1}{\mathcal C}.
\label{V/VA}
\end{align}
Accordingly, the condition (\ref{JWD-AB}) of Alice-to-Bob Gaussian unsteerability holds no matter 
whether Alice's party does or does not fulfill the requirement of quantumness 
\begin{align}
{\mathcal V_A}+\frac{i}{2} J_A \geq 0.
\label{quantVA}
\end{align}
Note that this aspect of steering of Bob's party by Alice's measurement is also at the heart 
of our UR treatment according to Eqs. (\ref{ABus}) and (\ref{usAB}).

The inequality (\ref{LSA}) implies a numerical necessary condition of Alice-to-Bob Gaussian unsteerability,
\begin{align}  
\frac{\det ({\mathcal V} )}{\det\left( {\mathcal V}_A \right)}  \geqq \frac{1}{2^{2M}},
\label{SchurAB}
\end{align}
which is similar to the condition (\ref{ncusBA}) for Bob-to-Alice unsteerability. Except for the case $M=1$,
the necessary condition (\ref{SchurAB}) is weaker than the matrix inequality (\ref{LSA}), so that it is not
a sufficient condition for Alice-to-Bob Gaussian unsteerability.

In this work our interest is restricted to bipartite $(N \,\text{vs} \,1)$-mode states.  By using suitable
EPR-like URs, we have derived the {\em necessary} matrix conditions (\ref{JWD-AB}) and (\ref{JWD-BA})      
for their one-way unsteerability, {\em irrespective} of their Gaussian or non-Gaussian nature. Concerning
the well-known sufficiency of these conditions for GSs and Gaussian measurements \cite{W1,W2}, 
we just take advantage of the above discussion to stress that the ratio 
${\det ({\mathcal V} )}/{\det\left( {\mathcal V}_A \right) }$ is a valuable indicator of Gaussian unsteerability 
from Alice to Bob, when compared to the number $1/4$, as pointed out by Eq. (\ref{ncusAB}).

\section{Summary and discussion} 

A main issue we are concerned with in the present work is to give a unified description 
of both entanglement and EPR steering for a class of bipartite multimode states 
of continuous-variable systems by using an efficient  theoretical method. Looking 
at the impressive volume of recent work on steering matters one can notice 
that the question on the relationship between criteria of entanglement and EPR steering 
is scarcely addressed. Accordingly, the present work on inseparability and steering conditions 
for two-party multimode quantum states is built on the idea of the fundamental role 
the sum-form URs of nonlocal EPR-like observables play in defining and describing 
these two types of quantum correlations. 

A widespread classification of quantum correlations termed as entanglement, EPR steering, 
and Bell nonlocality, from a quantum-information perspective was given by Wiseman {\em et al.} 
\cite{W1,W2,W3}.  Adopting here this fruitful point of view, we extend the quantum-mechanical
line of reasoning initiated by Reid for a two-mode state \cite{Reid1} by applying it to the whole class 
of two-party $(N \,\text{vs} \,1)$-mode states. A parallel analysis of their entanglement and steering 
makes use of the same pair of nonlocal EPR-like observables introduced in Eq. (\ref{QP}). 
Essentially, we take two steps. First, we employ the theorem of Hofmann and Takeuchi \cite{HT} 
to write the uncertainty relations with the sum of variances of the EPR-like observables (\ref{QP}) 
for separable states, Eq. (\ref{sepSUR}), and, respectively, for unsteerable ones,  Eqs. (\ref{usAB}) 
and (\ref{usBA}). 

The second step consists in looking for the extremal normalized sums of uncertainties with respect 
to the involved parameters $ \bm{\alpha},\bm{\beta}$. We start with the necessary conditions 
of separability (\ref{sepnc}). Although not exploited analytically for $N>1$, they exhibit  the PPT property, 
as expected from Peres' general theorem \cite{Peres}. Nevertheless, in the case of two-mode states 
$(N=1)$, we readily recover the minimum $2{\kappa}_{-}^{\rm PT}$ for $d<0$ \cite{PT2018}, 
as specified in Eqs. (\ref{minS1}) and (\ref{sgn(d)}).  

Then we focus on the steering scenario which, unlike entanglement, is known to be asymmetric 
with respect to the interchange of the two parties. The necessary condition of unsteerability 
from Alice to Bob, Eq. (\ref{ncusAB}), is the only numerical one to be equivalent to matrix inequalities 
having the form (\ref{ABusV}). By contrast, when $N>1$, the equivalent matrix inequalities 
(\ref{BAusV/VB}) and (\ref{BAusV}), requested by the Bob-to-Alice unsteerability, are stronger 
than the numerical one, Eq. (\ref{ncusBA}). However, the similar inequalities (\ref{ncusAB}) and
(\ref{ncusBA}) are valuable since their violation allows one to recognize the steerability 
from Alice to Bob and, respectively, from Bob to Alice.

We finally recall the privileged position of the two-party $(N \,\text{vs} \,1)$-mode GSs, 
for whom the above necessary conditions regarding both kinds of quantum correlations 
are equally sufficient ones. We start with Simon's seminal work \cite{Simon}, who essentially
proved that the first inequality (\ref{sepnc1}), ${\kappa}_{-}^{\rm PT}  \geqq 1/2,$  
is a criterium of separability for two-mode GSs. The result of Werner and Wolf \cite{WW2001} 
that the PPT property of {\em any} $(N \, \text{vs} \,1)$-mode GS implies its separability 
is the maximal generalization of Simon's theorem for $N=1$ \cite{Simon}. 
However, by contrast to the latter, it provides no analytic separability indicator for $N>1$.

Then we apply the unsteerability criterion for two-party $(N \,\text{vs} \, M)$-mode GSs, 
under Gaussian measurements, found by Wiseman {\em et al.} in Refs. \cite{W1,W2}.
Accordingly, the matrix inequalities (\ref{JWD-AB}) and (\ref{JWD-BA}), written for a bipartite
$(N \,\text{vs} \,1)$-mode GS ${\hat \rho}_G$ shared by Alice and Bob,  are necessary and 
sufficient conditions of its unsteerability from Alice to Bob and, respectively, from Bob to Alice.
We emphasize that the necessity part is proven here with no reference to the Gaussian or
non-Gaussian nature of the state, as well as to that of the one-party measurements involved.

Let us make a comment about three sets of $(N \,\text{vs} \,1)$-mode states having 
the {\em same} CM ${\mathcal V}$ in the standard form (\ref{Vq+Vp}): the set ${\cal S}$ 
of separable states, as well as the sets ${\cal U}^{(A \to B)}$ and ${\cal U}^{(B \to A)}$ 
of unsteerable states from Alice to Bob and from Bob to Alice, respectively. All the states 
belonging to ${\cal S}$ fulfill the EPR-like necessary conditions of separability (\ref{sepnc}), 
while those belonging to the sets ${\cal U}^{(A \to B)}$ and  ${\cal U}^{(B \to A)}$ satisfy 
the necessary matrix conditions of unsteerability (\ref{ABusV}) and, respectively, (\ref{BAusV}).
However, by virtue of the Werner-Wolf theorem \cite{WW2001}, for the GSs in ${\cal S}$, 
defined up to a phase-space translation, the conditions of separability (\ref{sepnc}) are, 
in addition, {\em sufficient} ones. Similarly, in view of a theorem proven by Wiseman {\em et al.} 
\cite{W2}, for the GSs belonging to the sets ${\cal U}^{(A \to B)}$ and  ${\cal U}^{(B \to A)}$ 
and under one-party Gaussian measurements, the matrix conditions of unsteerability 
(\ref{ABusV}) and (\ref{BAusV}) are also {\em sufficient} ones. These remarks  are just
some aspects of the special role played by the GSs among all $(N \,\text{vs} \,1)$-mode 
states with the same CM. The best known example of extremal property of the GSs
within the multimode states with a given CM is that they attain the maximum 
von Neumann entropy \cite{Holevo1999}. Other relevant examples are mentioned 
and discussed in Ref. \cite{extrem}.

We finally emphasize that we have developed in this work a unified treatment 
of inseparability and steering of two-party $(N \,\text{vs} \,1)$-mode states,
based on specific sum-type URs of suitable EPR-like observables. Extremization 
of the associate normalized sums of variances provides the right necessary conditions
for separability and one-way unsteerability. In the Gaussian framework, these 
conditions are, at the same time, the well-known sufficient ones.

\section*{Acknowledgment}

This work was supported by the Funding Agency CNCS-UEFISCDI of the Romanian Ministry of Research 
and Innovation through grant No. PN-III-P4-ID-PCE-2016-0794.

\end{document}